\documentclass[onecolumn,showpacs,preprintnumbers,amsmath,amssymb,superscriptaddress]{revtex4}

\usepackage{graphicx}
\usepackage{dcolumn}
\usepackage{bm}
\usepackage{longtable}

\begin{document}
\title{Diffusive thermal dynamics for the spin-S Ising ferromagnet}

\author{E. Agliari}
\affiliation{Dipartimento di Fisica, Universit\`a degli Studi di Parma, Viale Usberti 7/a, 43100 Parma, Italy} 
\author{R. Burioni}
\affiliation{Dipartimento di Fisica, Universit\`a degli Studi di Parma, Viale Usberti 7/a, 43100 Parma, Italy}
\author{D. Cassi}
\affiliation{Dipartimento di Fisica, Universit\`a degli Studi di Parma, Viale Usberti 7/a, 43100 Parma, Italy}
\author{A. Vezzani}
\affiliation{CNR- INFM Gruppo Collegato di Parma, Viale Usberti 7/a,
43100 Parma, Italy}
%

%
\date{\today}
%
\begin{abstract}
We introduce an alternative thermal diffusive dynamics
for the spin-S Ising ferromagnet realized by means of a random
walker. The latter hops across the sites of the lattice and flips
the relevant spins according to a probability depending on both
the local magnetic arrangement and the temperature. The random
walker, intended to model a diffusing excitation, interacts with
the lattice so that it is biased towards those sites where it can
achieve an energy gain. In order to adapt our algorithm to systems
made up of arbitrary spins, some non trivial generalizations are
implied. In particular, we will apply the new dynamics to
two-dimensional spin-1/2 and spin-1 systems analyzing their
relaxation and critical behavior. Some interesting differences
with respect to canonical results are found; moreover, by
comparing the outcomes from the examined cases, we will point out
their main features, possibly extending the results to spin-S
systems.
\end{abstract}
\maketitle
\section{Introduction}
\label{intro} The Ising model has been extensively studied both by
analytical and computational methods; the latter are especially
useful for complex and high dimensional lattices and rely, for
example, on Monte Carlo methods \cite{deng}. This implies to find
a prescription for updating the spin system and an algorithm which
determines if the suggested spin-flip can be accepted.

The first procedure is the most subtle and it is usually chosen so
that it can be easily implemented (it typically consists in a
sweep along parallel lattice lines) while the latter often refers
to well-known algorithms such as the Glauber one.

Here our aim is not to find an efficient algorithm, but rather to
realize a thermal dynamics physically consistent, which possibly
violates the detailed balance condition. In particular, we refer
to \cite{buonsante} where a diffusive dynamics was introduced: the
spin flips are induced by a random walker hopping across the sites
of the lattice. This model was inspired by some
non-stechiometrical compounds \cite{dagotto} where diffusing
excitations (for example charged carriers) affect the spin
dynamics. Then, in our dynamics, the walker is meant as a local
excitation diffusing throughout the whole sample and interacting
with the magnetic arrangement. Moreover, we suppose the walker to
be biased towards those sites where a spin-flip is energetically
more favorable. This technique is not only more natural than the
traditional ones, but it can also be applied to irregular lattices
represented by graphs.

The previous work succeeded in defining a new, well-working
dynamics, nevertheless the algorithm introduced was expressly
meant for a spin-1/2 system. Its extension to the general spin-S
case is non trivial since, while in the spin-1/2 case, each spin
of the lattice allows only one possible new state, here the spin
status is not binary and a manifold choice occurs. Therefore, a
further random process has to be introduced: apart from the one
concerning the selection of the nearest-neighbor to move towards,
we also have to take into account the one relevant to the variety
of states accessible to the spin considered. Then, in this work,
we developed a new algorithm able to be applied to systems made up
of discrete spins with an arbitrary number of states. Not only, we
also wondered to what extent results found in \cite{buonsante}
depend on the special algorithm and spin model taken into account.
In order to do so we implemented our dynamics on both spin-1/2 (as
a test) and spin-1 (as first example) systems.

The remaining of the paper is organized as follows. In
Secs.~\ref{model} and \ref{dtd} we explain the model and the new
algorithm, Sec.~\ref{thermo} and \ref{relaxation} are devoted to
the analysis of the results: thermodynamics of the system and
relaxation at low temperature, respectively. Finally, in
Sec.~\ref{concl} we discuss our outcomes.

\section{The model}
\label{model} The most general spin-1 Ising model with up-down
symmetry is the BEG model \cite{beg}, whose Hamiltonian reads:
\begin{equation}
\label{eq:hamiltonian} {\cal H} = -J\sum_{j\sim i}^{N} \sigma_i
\sigma_j - D\sum_{j\sim i}^{N} \sigma_i^2 \sigma_j^2 -
K\sum_{i}^{N} \sigma_i^2,
\end{equation}
where the first two sums are over all nearest-neighbor pairs on
the lattice, the last is over all sites and $\sigma_{i}=\pm $1,0.
This model was originally introduced to study phase separation and
superfluidity in $^{3}$He - $^{4}$He mixtures; then it has been
applied to describe properties of multicomponent fluids,
microemulsions, superconductor alloys and electronic conduction
models \cite{lajzerowicz,schick,newman1}. Here we consider the
particular situation with K=D=0 and $J\ne 0$ in order to preserve
the analogy with the spin-1/2 case and to concentrate on the
dynamical aspects.

The analysis of the diffusive dynamics is carried out from the
numerical point of view adopting a two dimensional array of spins,
so that, as mentioned above, results obtained for the spin-1/2 are
useful as a test by comparison with those analytically known and
relevant to the canonical equilibrium state. Unfortunately, this
is not possible for the spin-1 case as there exists no exact
solution, hence we will refer to earlier works mainly dealing with
Monte Carlo simulations, finite-size scaling, high- and
low-temperature expansions
\cite{hoston,berker,adler,wang,dasilva,hont}.

However, just a relatively small number of works about critical
exponents for the $S \geq 1$ Ising model has been published. Those
works are mostly numerical and they confirm the exponents
independence on the spin magnitude, as consistent with the
renormalization group theory \cite{jensen}. Therefore, our work,
though based on a non-traditional dynamics, would offer an insight
into this matter. In fact, also encouraged by the interesting
outcomes found in \cite{buonsante}, we meanly focused on the
critical aspects.

\section{\label{dtd}Diffusive Thermal Dynamics\protect\\}

We refer to the algorithm introduced in \cite{buonsante} and we
improve it so that it can be easily adapted to systems made up of
spins with an arbitrary number of states $q$. In fact, as already
mentioned, that kind of algorithm is an exclusive for systems made
up of binary valued spins.

The relaxation dynamics is realized by a random walker diffusing
through the sites of the Ising lattice. In general, the walker on
a site $i$ has $(2d+1)q$ possibilities: it can move towards one of
its $2d$ nearest neighbors $j$ or stop and it can flip the spin
relevant to the reached site or leave it unchanged.

More precisely, the walker moves from $i$ to $j$ realizing the
magnetic configuration $\vec{s}$ according to the normalized
probability:
\begin{equation} \label{eq:probability}
{\cal P}_T(\vec{s},i,j) =
\frac{p_T(\vec{s},j)}{\sum_{\{\vec{s'}\}} \sum_{j=0}^{2d}
p_T(\vec{s'},j)}.
\end{equation}
In this equation $\{\vec{s'}\}$ is the whole of magnetic
configurations which can be realized from the current one and
\begin{equation} \label{eq:Glauber_probability}
p_T(\vec{s},k)=\frac{1}{1+e^{[\beta \Delta E_k(\vec{s})]}}
\end{equation}
represents the probability of spin-flip relevant to the site $k$,
being
\begin{equation} \label{eq:delta_energy}
\Delta E_k(\vec{s}) = (\sigma_k'-\sigma_k)\sum_{j\sim i}\sigma_i,
\end{equation}
the energy variation consequent to the process.
Eq.~(\ref{eq:Glauber_probability}) has been derived from the usual
Glauber probability \cite{newman2}:
\begin{equation} \label{eq:probabilityG}
P_T^G(\vec{s},k) = \frac{e^{- \beta E_n}}{\sum_{m=1}^{q} e^{-
\beta E_m}}
\end{equation}
which represents the probability that the selected spin $k$ has
the value $n$ ($1 \leq n \leq q$), where $E_n$ is the energy of
the system when $\sigma_k = n$. Note that the previous expression
can be rewritten as
\begin{equation}
P_T^G(\vec{s},k) = \frac{1}{\displaystyle 1 + \sum_{m=1 \atop m
\neq n }^{q} e^{\beta \Delta E_m}}
\end{equation}
with $\Delta E_m = E_n - E_m$ and it reduces to
Eq.~(\ref{eq:Glauber_probability}) when $q = 2$. However, we adopt
Eq.~(\ref{eq:Glauber_probability}) in each case because it is more
direct and it also reveals to be more efficient.

You can notice that the magnetic configuration of the system, as
well as the position of the walker, can remain unchanged and that
the diffusion of the walker is biased towards those sites where it
can achieve a gain in the energy.

There are some important consequences of the fact that such a
dynamics includes both the walker motion on the lattice and the
magnetic evolution of the lattice itself. In particular, the
analytical approach is made rather difficult and the detailed
balance is explicitly violated. In fact, the latter imposes the
quite restrictive condition:
\begin{equation} \label{eq:detailed}
p_{\nu} P(\nu \rightarrow \mu) = p_{\mu} P(\mu \rightarrow \nu),
\end{equation}
according to which the overall rate at which transitions from one
state $\nu$ to another state $\mu$ happen is the same for the
reverse process. However, in our system, the probability of being
in a state $\nu$, as well as the probability of making a
transition $\nu \rightarrow \mu$, are non trivial functions of
both the magnetic arrangement and the position of the walker on
the lattice, which involves that Eq.~(\ref{eq:detailed}) does not
hold. In order to clarify this subtle point, a further insight is
provided. Suppose the transition $\mu \rightarrow \nu$ represents
the walker jumping from site $i$ to $j$, realizing the spin-flip
$\sigma_j \rightarrow \sigma_j^{\ast}$. The reverse transition is
obviously impossible, since it requires the walker to flip the
spin relevant to the starting site, $\sigma_j^{\ast} \rightarrow
\sigma_j$, while jumping from $j$ to $i$. As mentioned at the
beginning of this section, this kind of flip is forbidden by our
dynamics so that $P(\nu \rightarrow \mu)=0$. On the other hand,
since the walker can reach any lattice site, whatever the magnetic
arrangement, both $p_{\mu}$ and $p_{\nu}$ are strictly positive
quantities; as a result Eq.~(\ref{eq:detailed}) is false.

Note that the violation of the detailed balance is consistent with
our dynamics intent: it is not meant to recover the canonical
distribution, but rather to model some possible physical processes
making the spin system evolve.

Analogous considerations can be drawn for other kinds of diffusive
dynamics employing random walkers.

\begin{figure}[tb]
\resizebox{0.495\columnwidth}{!}{\includegraphics{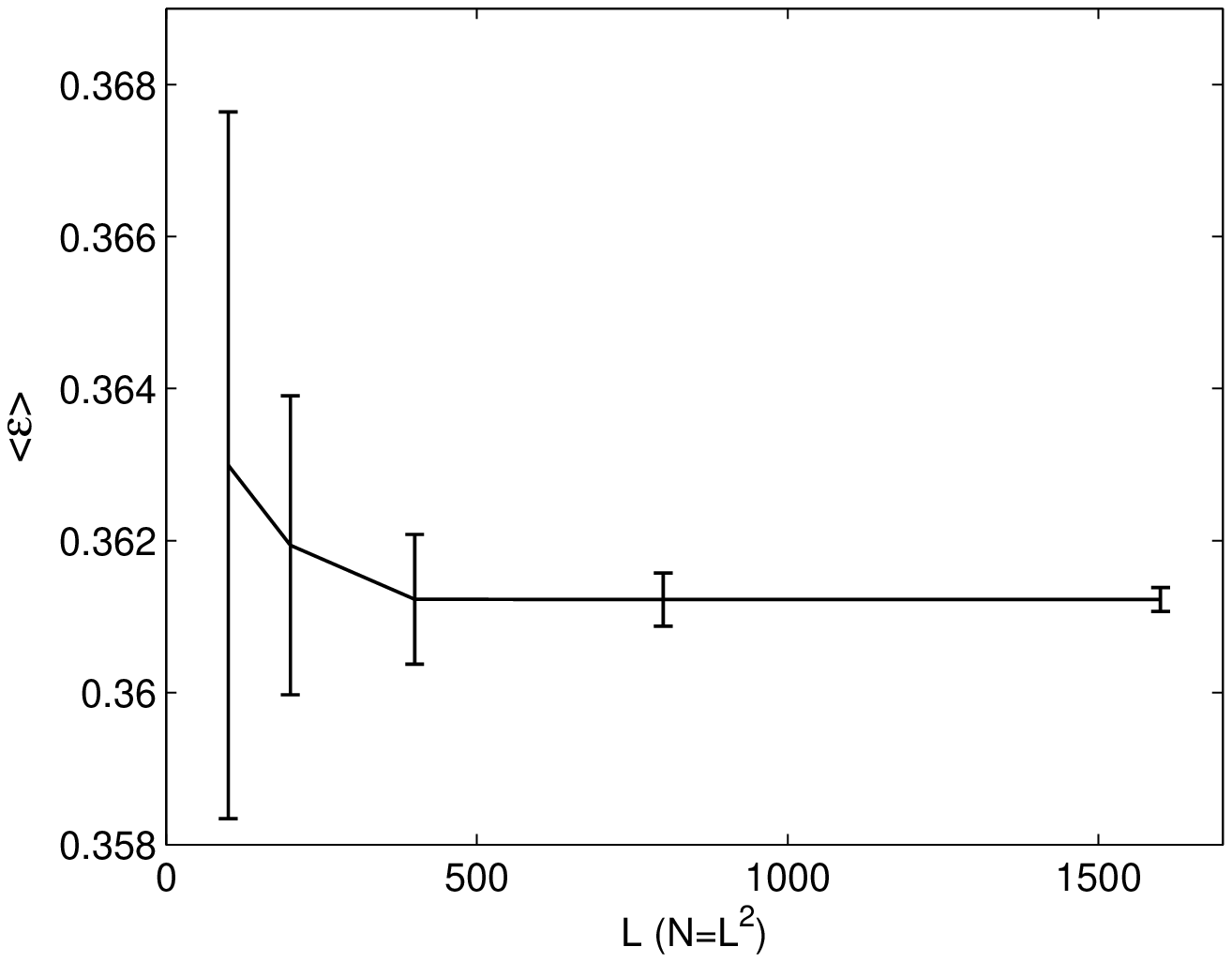}}
\resizebox{0.495\columnwidth}{!}{\includegraphics{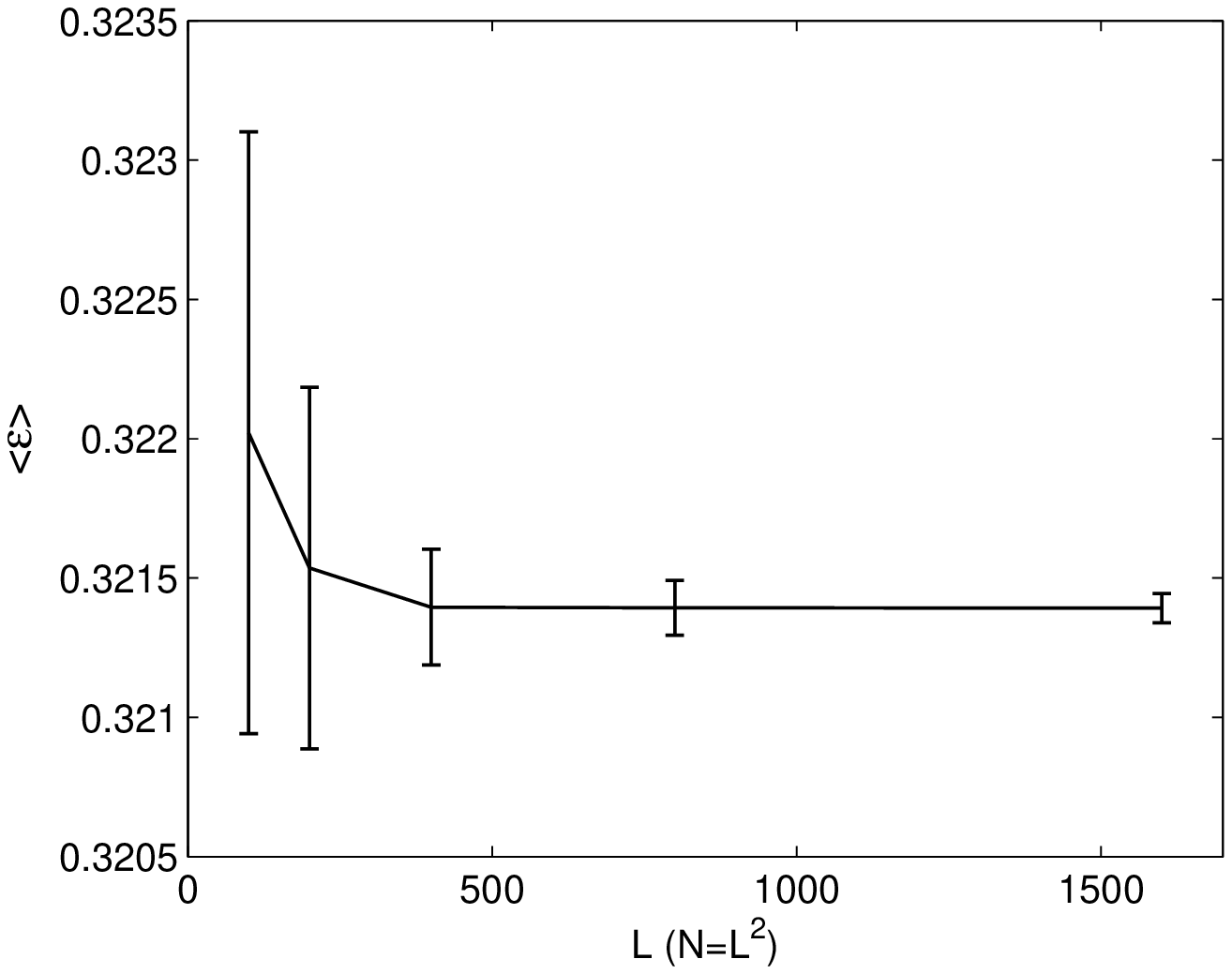}}
\caption{Finite Size scaling for the specific energy of a spin-1/2
(left panel) and spin-1 (right panel) Ising system subject to the
diffusive dynamics described in Sec.~\ref{dtd} at $T = 2.4$ and $T
= 1.56$ respectively. All the measurements were carried out in the
stationary regime and the error bars represent the fluctuations
about the average values.} \label{fig:fss_energy}
\end{figure}
\begin{figure}[tb]
\resizebox{0.495\columnwidth}{!}{\includegraphics{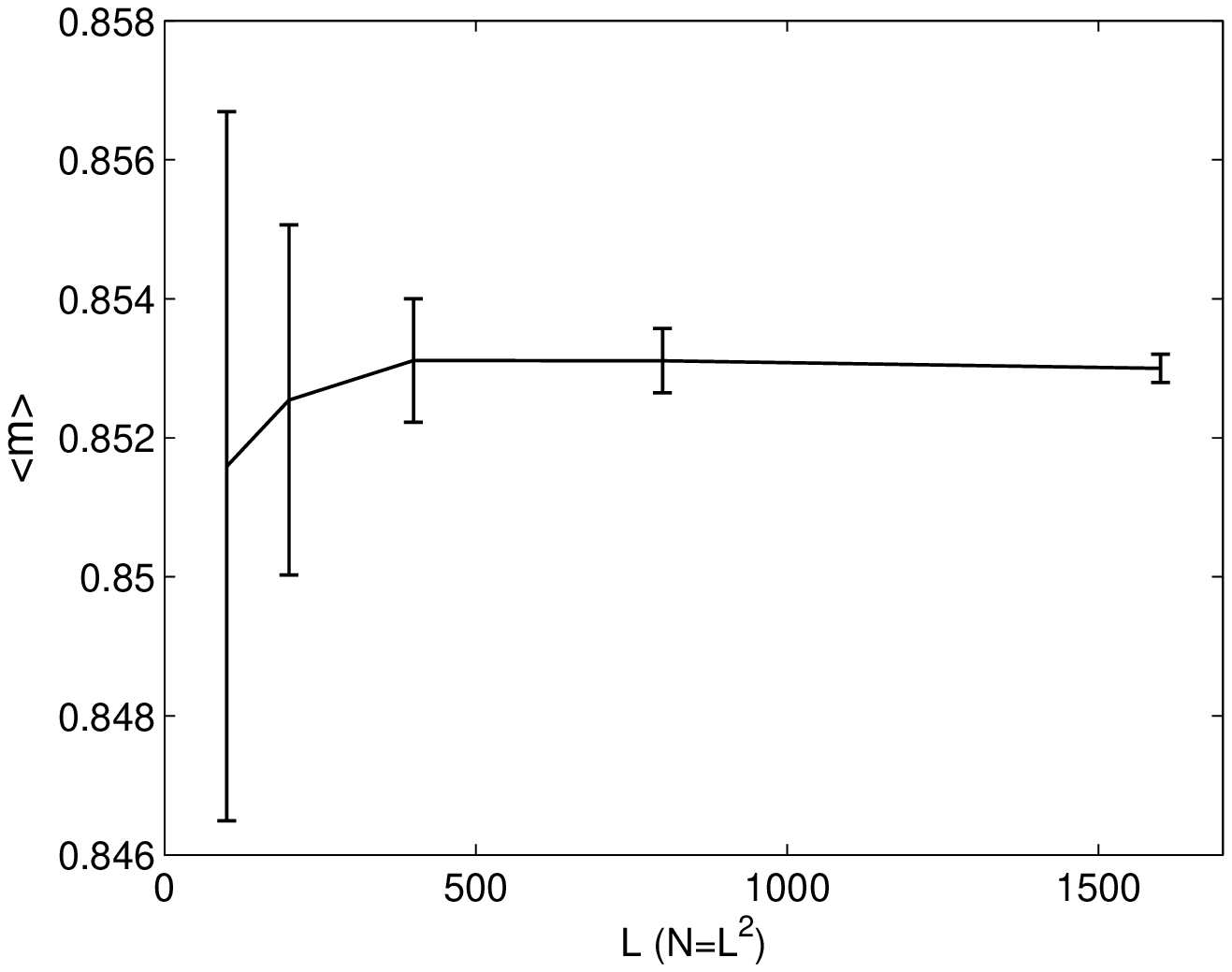}}
\resizebox{0.495\columnwidth}{!}{\includegraphics{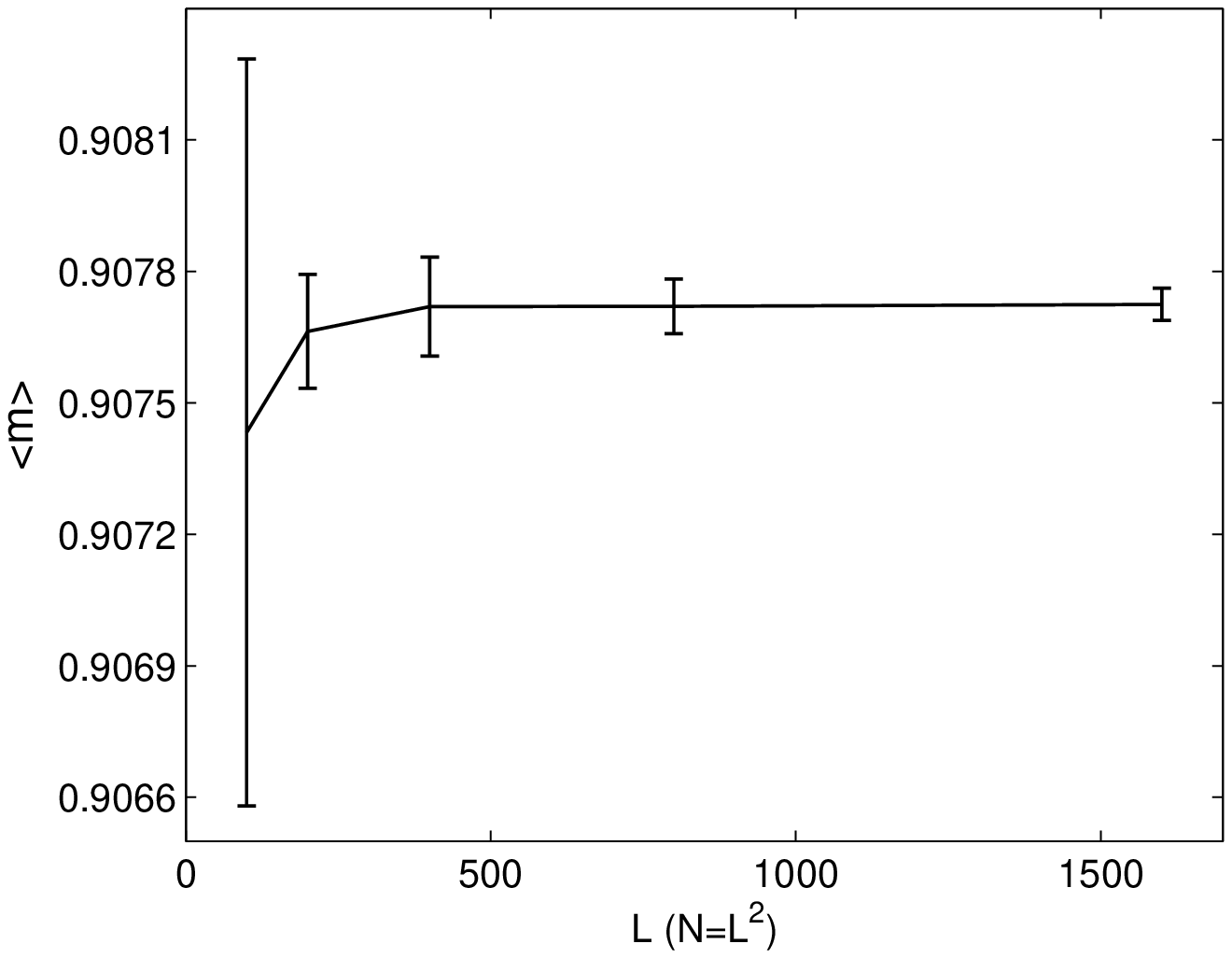}}
\caption{\label{fig:fss_mag}Finite Size scaling for the specific
magnetization of a spin-1/2 (left panel) and spin-1 (right panel)
Ising system subject to the diffusive dynamics described in
Sec.~\ref{dtd} at $T = 2.4$ and $T = 1.56$ respectively. All the
measurements were carried out in the stationary regime and the
error bars represent the fluctuations about the average values.}
\end{figure}

On the other hand, a fundamental difference with the algorithm
suggested in \cite{buonsante} is that here, once the new site
selected, the corresponding probability is not determined because
we have also to specify the magnetic configuration $\vec{s}$
candidate to be realized. Of course, in the spin-1/2 case this is
not necessary because, once the new site chosen, there is just one
new magnetic configuration which can be considered.

Thus, our algorithm generalizes the previous one: now it includes
all kinds of new scenarios so that it can properly work also for
$S > 1/2$ cases.

Our analysis will be performed mainly by means of numerical
simulations keeping fixed the value of the exchange interaction
constant (J=1) and setting periodic boundary conditions for the
square lattice where spins are placed on. In fact, it is clear
that an equilibrium situation can be reached only after the random
walker realizing the dynamics has visited every sites of the
system a sufficient number of times; this in particular selects
the periodic boundary conditions as the most natural for the
problem. Moreover, we deal with just one walker postponing the
case of a larger density to next works.

\section{Thermodynamic of spin-1/2 and spin-1 systems}
\label{thermo}

In this section we describe the results pertaining to spin-1/2
showing their consistency with those in \cite{buonsante} and then
we will move to spin-1 delaying a global discussion to
Sec.~\ref{concl}.
\begin{figure}[tb]
\resizebox{0.495\columnwidth}{!}{\includegraphics{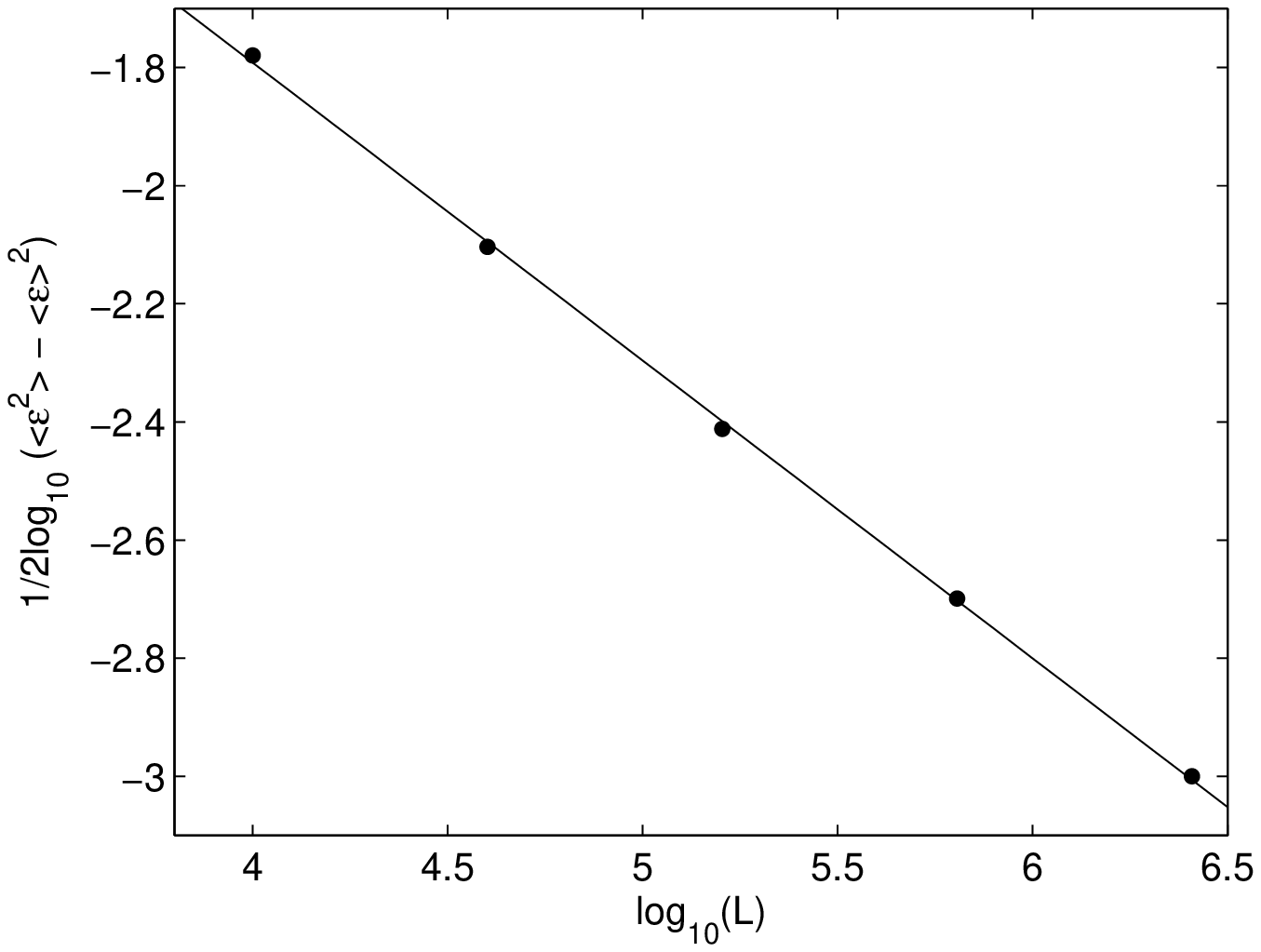}}
\resizebox{0.495\columnwidth}{!}{\includegraphics{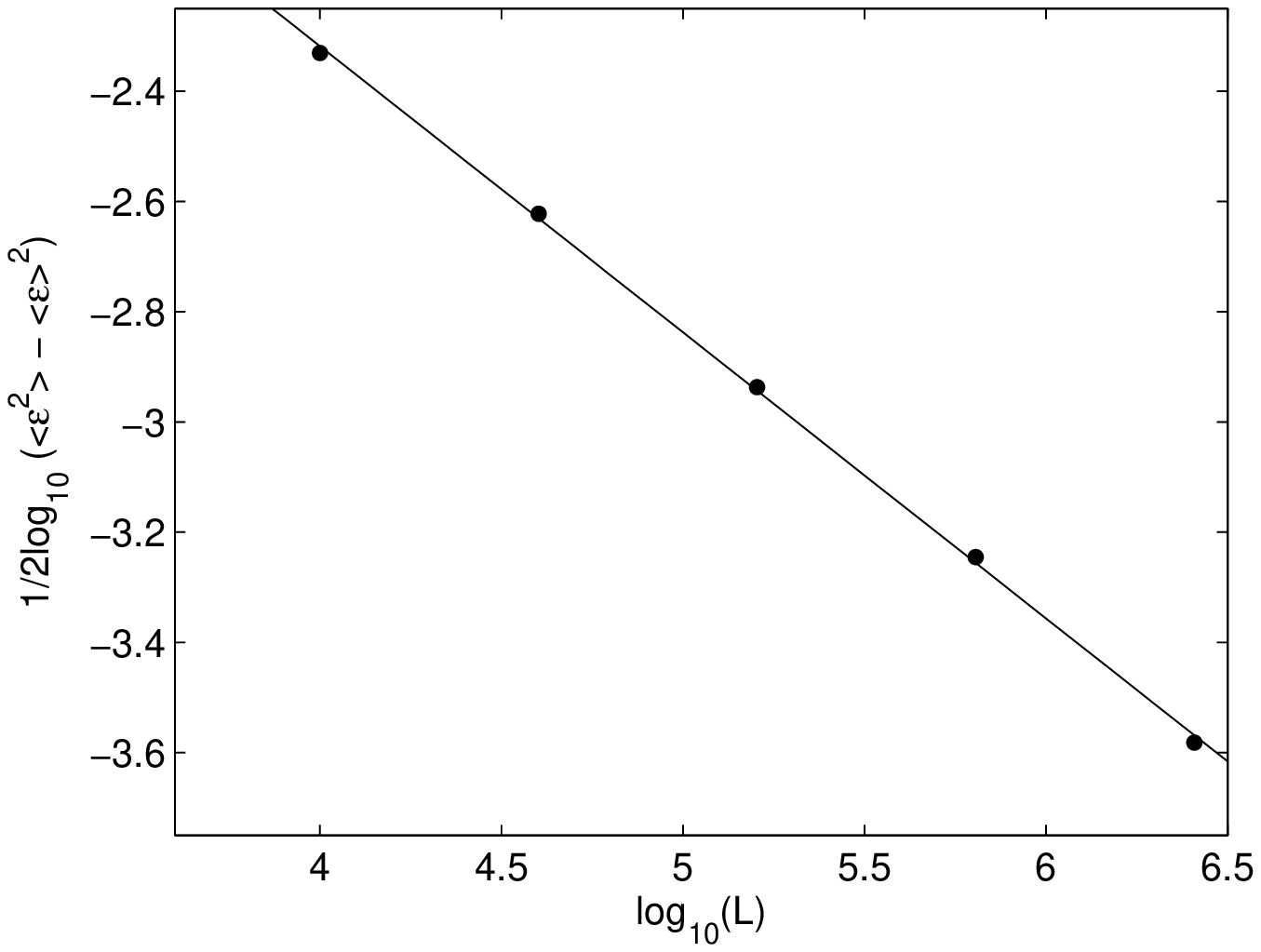}}
\caption{\label{fig:flutt_e}Finite Size scaling for the
fluctuation about the average value of the specific energy for a
spin-1/2 (left panel) and spin-1 (right panel) Ising system
subject to the diffusive dynamics described in Sec.~\ref{dtd} at
$T = 2.4$ and $T = 1.56$ respectively. The slopes of the linear
fit (line) of the measured data (\textbullet) $-0.50 \pm 0.01$ and
$-0.51 \pm 0.02$ are in good agreement with the expected value
$-0.5$. All the measurements were carried out in the stationary
regime.}
\end{figure}
In both cases the dynamics realized by the random walker actually
drives the system to a thermodynamically well-behaved steady
state, highly independent on the initial conditions. As we will
see later, results relevant to the critical exponents will provide
another strong signature that the stationary state reached by the
system is actually an equilibrium state, though it is non
trivially different from the canonical equilibrium of the Ising
model. In fact, we verified that also for our diffusive dynamics
the normalized joint probability $\widetilde{{\cal
P}}(\epsilon,m,T)$, introduced in \cite{buonsante}, depends on
$T$.

\begin{figure}[tb]
\resizebox{0.495\columnwidth}{!}{\includegraphics{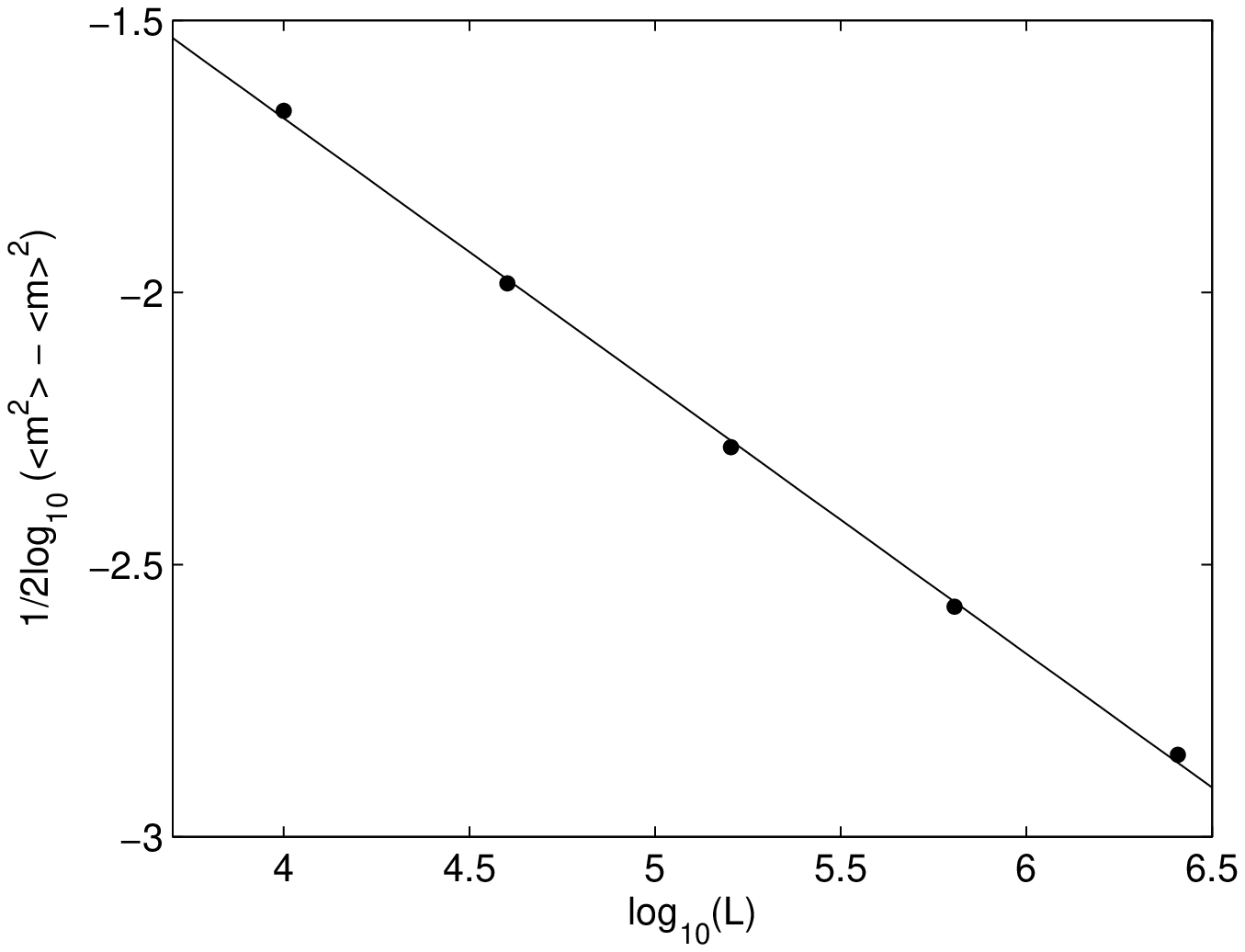}}
\resizebox{0.495\columnwidth}{!}{\includegraphics{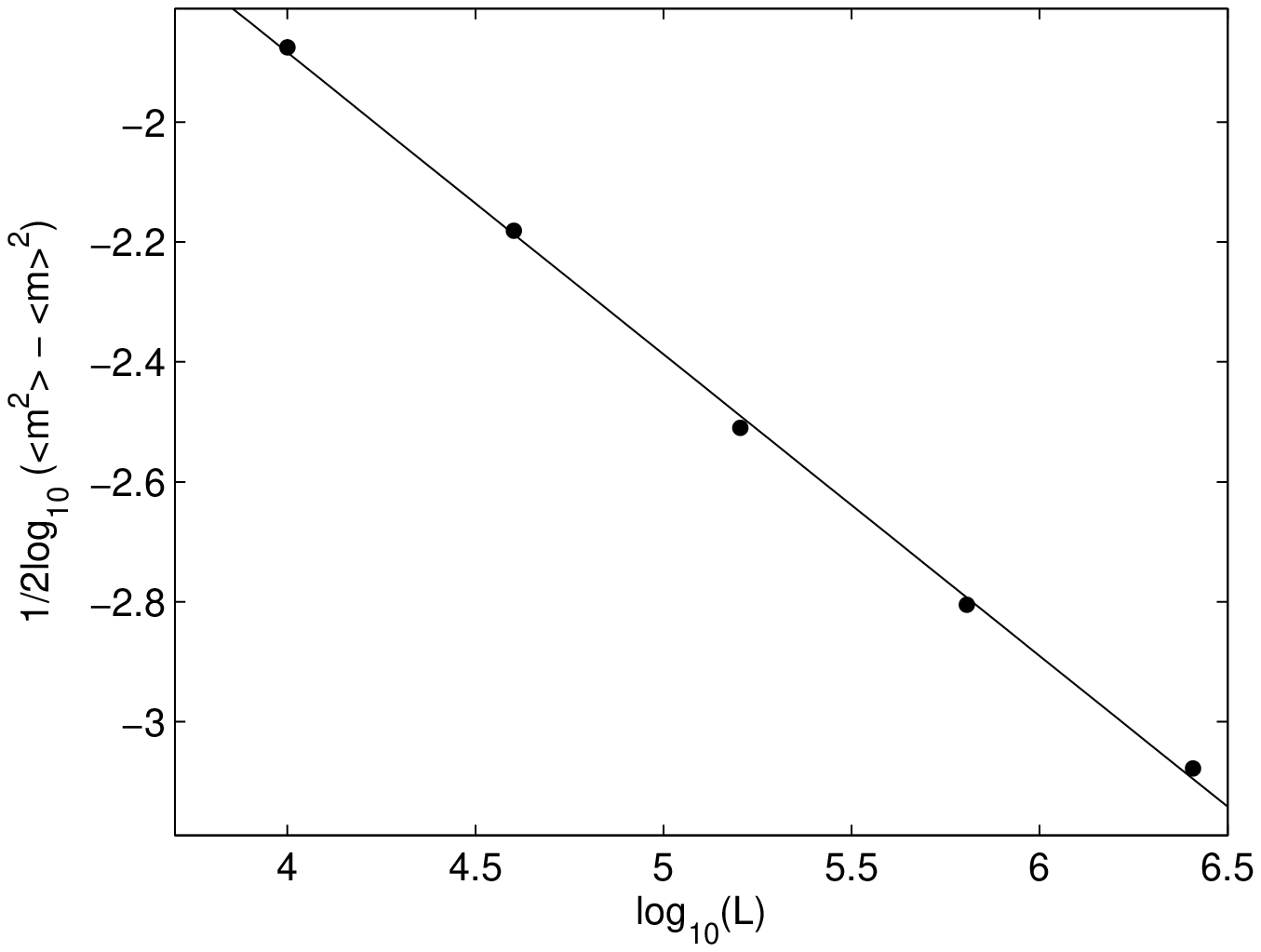}}
\caption{\label{fig:flutt_m}Finite Size scaling for the
fluctuation about the average value of the specific magnetization
for a spin-1/2 (left panel) and spin-1 (right panel) Ising system
subject to the diffusive dynamics described in Sec.~\ref{dtd} at
$T = 2.4$ and $T = 1.56$ respectively. The slopes of the linear
fit (line) of the measured data (\textbullet) $-0.49 \pm 0.02$ and
$-0.50 \pm 0.02$ are in good agreement with the expected value
$-0.5$. All the measurements were carried out in the stationary
regime.}
\end{figure}

In Figs.~\ref{fig:fss_energy} and \ref{fig:fss_mag} the average
values of the specific energy and magnetization are plotted for
systems with different sizes, at a fixed value of the temperature
parameter. Figs.~\ref{fig:flutt_e} and \ref{fig:flutt_m} show the
expected scaling behavior for the fluctuations about the average
value of the specific thermodynamic observables which decrease as
the inverse square root of the lattice size.

\begin{figure}[tb]
\resizebox{0.6\columnwidth}{!}{\includegraphics{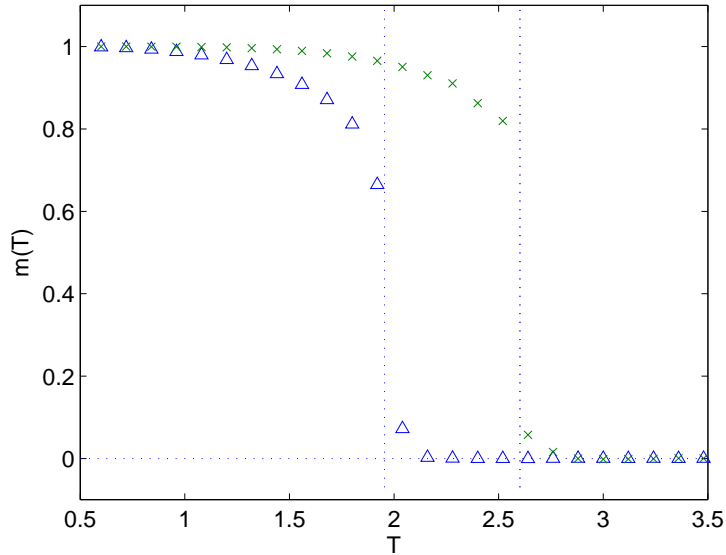}}
\caption{\label{fig:termo_mag} Specific magnetization for a $400
\times 400$ spin-1/2 ($\times$) and spin-1 ($\triangle$) Ising
system. The vertical dashed lines are placed at the critical
values of the temperature.}
\end{figure}

\begin{figure}[tb]
\resizebox{0.6\columnwidth}{!}{\includegraphics{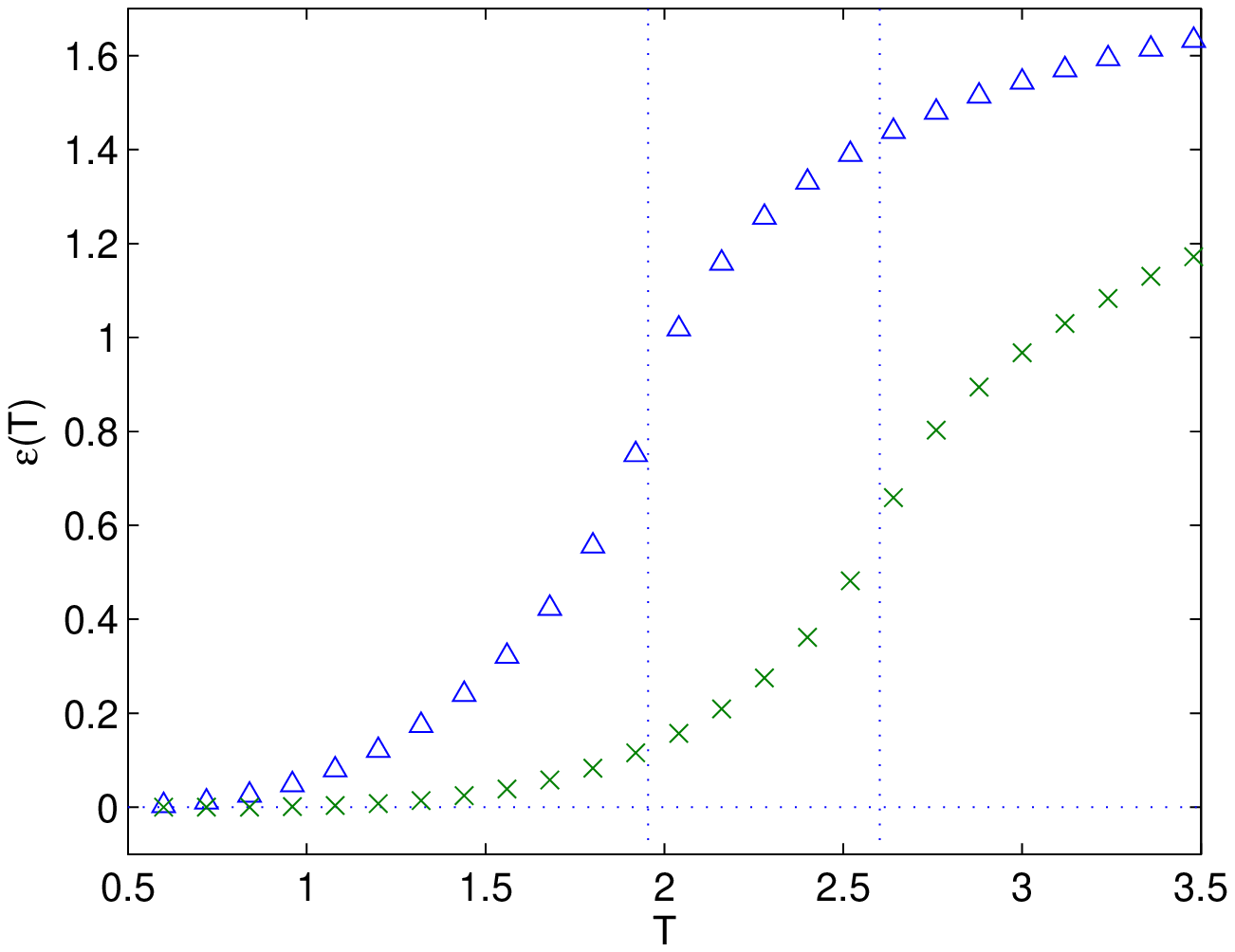}}
\caption{\label{fig:termo_e} Specific energy for a $400 \times
400$ spin-1/2 ($\times$) and spin-1 ($\triangle$) Ising system.
The vertical dashed lines are placed at the critical values of the
temperature.}
\end{figure}

\begin{figure}[tb]
\resizebox{0.45\columnwidth}{!}{\includegraphics{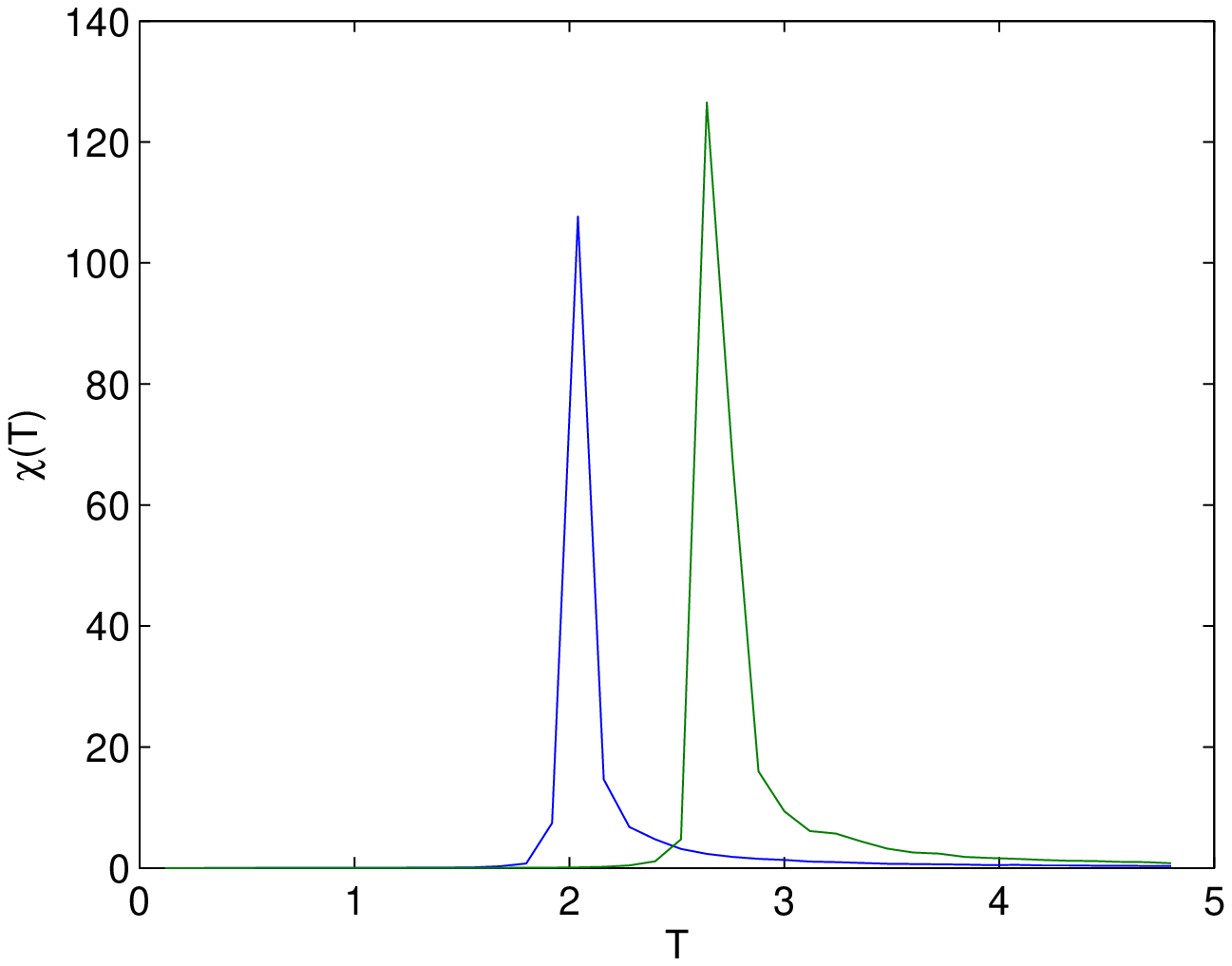}}
\resizebox{0.45\columnwidth}{!}{\includegraphics{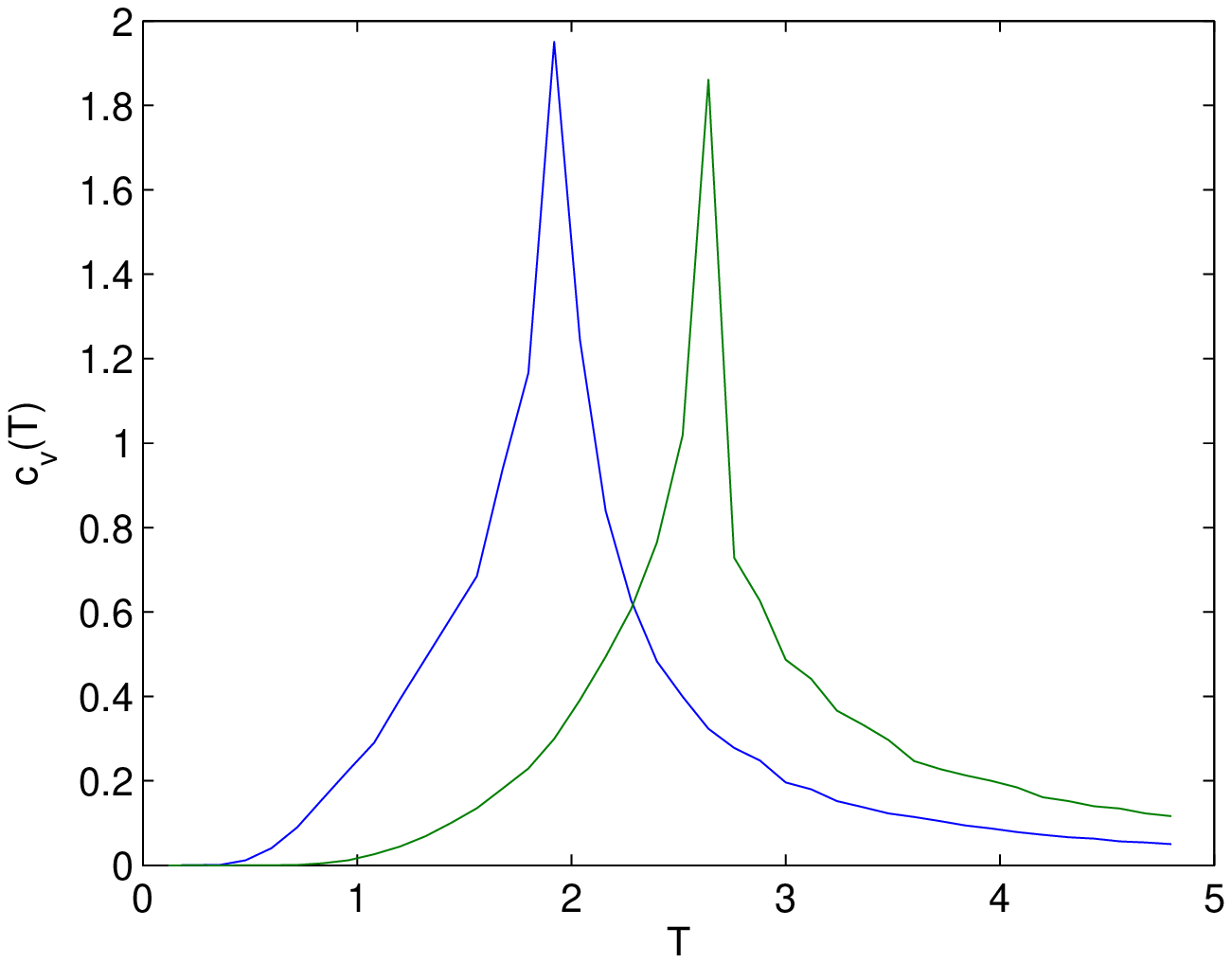}}
\caption{\label{fig:termo2_Mezzo} Magnetic susceptibility (left
panel) and specific heat (right panel) for a $400 \times 400$
Ising system of spin-1/2 and spin-1. The two cases are easily
distinguishable since the former displays a higher critical
temperature.}
\end{figure}

Now let us consider Figs.~\ref{fig:termo_mag} and
\ref{fig:termo_e}: the average values of magnetization and energy
are plotted versus temperature. For spin-1/2 a phase transition is
apparent at about $T = 2.6$ which is a value significantly higher
than the exact critical temperature (we will deeply return on this
feature later). For spin-1 we see similar, but somehow
left-shifted, curves which clearly suggest
$T_{c}^{S=1}<T_{c}^{S=1/2}$.
Analogous considerations can be made from
Fig.~\ref{fig:termo2_Mezzo} where relevant magnetic susceptibility
and specific heat are depicted: their profiles are consistent with
the theory and highlight that a phase transition happens at a well
defined temperature.
Note that these results do not depend on the particular initial
configuration which can, at least, affect the orientation of the
asymptotic arrangement. The evolution of the system when the
initial magnetization is very low is quite interesting, especially
in the spin-1 case, and it will be treated in the next Section.

Now we focus our attention on the critical behavior of the
systems, i.e. the properties featured nearby the phase transition.
From general theoretical considerations, based on the
renormalization group theory, we expect that the critical
exponents do not depend on the spin magnitude, but they are
characterized by the dimensionality of the system and by its order
parameter \cite{jensen}. Nevertheless, it is not trivial that a
diffusive dynamics, generating a non canonical ensamble, does not
affect the universality class.

First of all, we observe that, like for the canonical Ising model,
the phase transition induced by the diffusive dynamics exhibits a
singular behavior for the thermodynamic functions. In this context
it is important to stress that specific heat and magnetic
susceptibility were calculated as fluctuations according to other
studies of Ising system where fluctuation-dissipation theorem does
not strictly apply.


\begin{figure}[tb]
\resizebox{0.6\columnwidth}{!}{\includegraphics{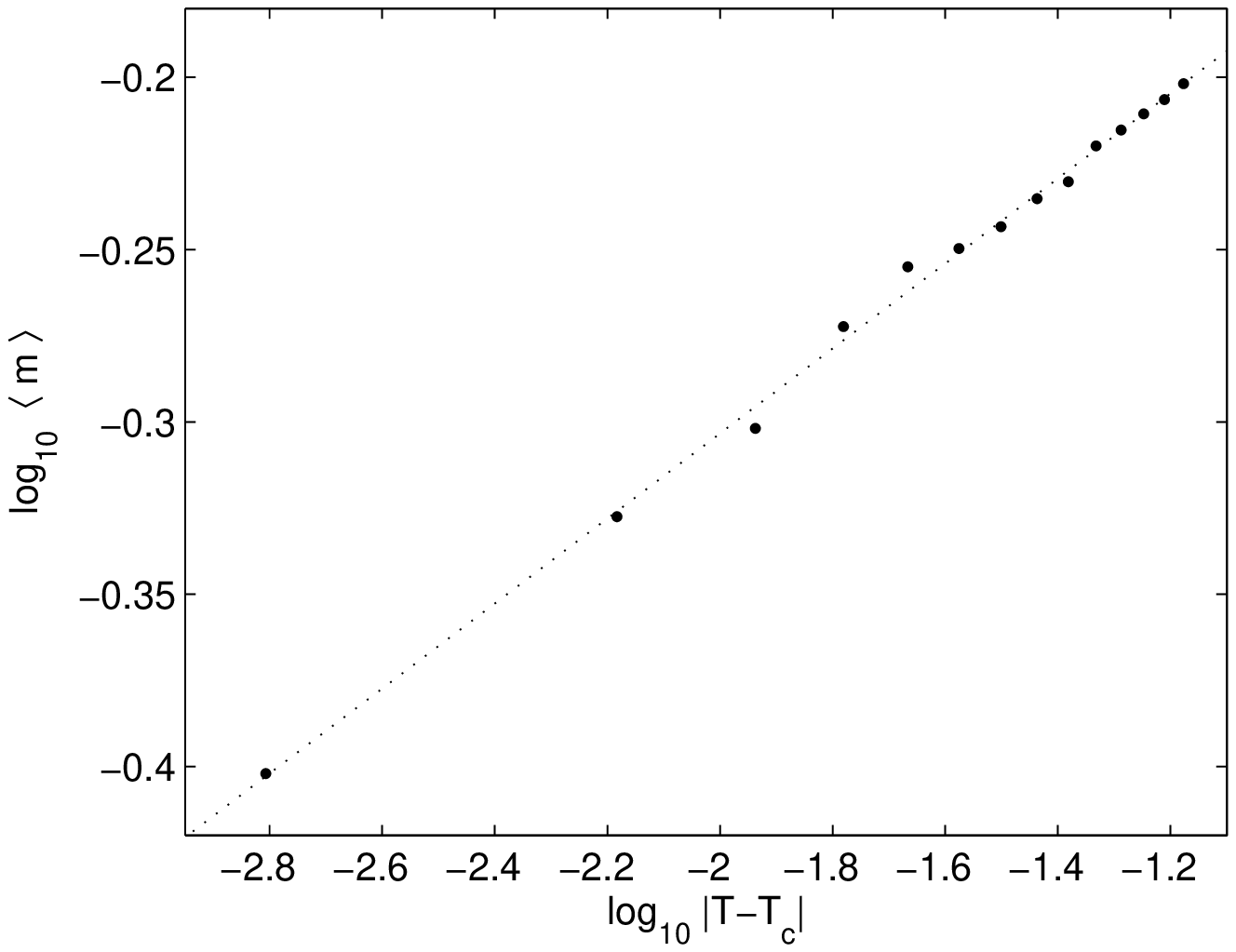}}
\caption{\label{fig:mcrit_Mezzo} Log-log scale plot of
magnetization versus $|T-T_c|$ for a spin-1/2 Ising system subject
to the diffusive dynamics ($\bullet$). The measures were performed
on a $1600 \times 1600$ array of spins. The dotted line is the
best fit: $y = A |T-T_c|^{\beta}$. The estimated values for the
critical temperature and for the exponent are $T_c = 2.602 \pm
0.001$ and $\beta = 0.123 \pm 0.005$, respectively. The latter is
consistent with the relevant canonical one.}
\end{figure}

\begin{figure}[tb]
\resizebox{0.6\columnwidth}{!}{\includegraphics{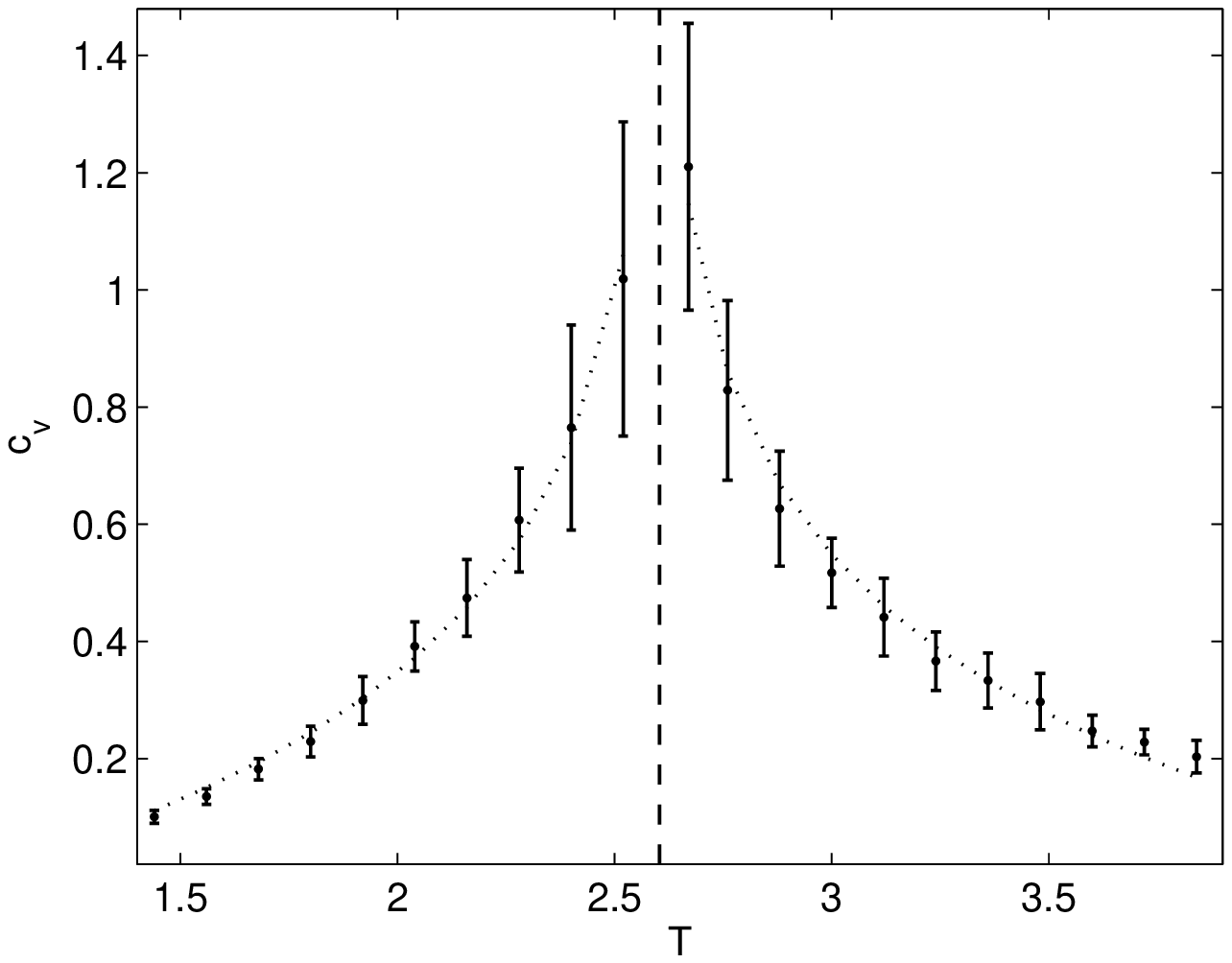}}
\caption{\label{fig:cvcrit_Mezzo} Specific heat for a spin-1/2
Ising system subject to the diffusive dynamics ($\bullet$). The
dotted curves fitting the data are of the form $f(T) = a + b ~
log(|T-T_c|)$. The vertical dashed line indicates the estimated
value of the critical temperature.}
\end{figure}

\begin{figure}[tb]
\resizebox{0.6\columnwidth}{!}{\includegraphics{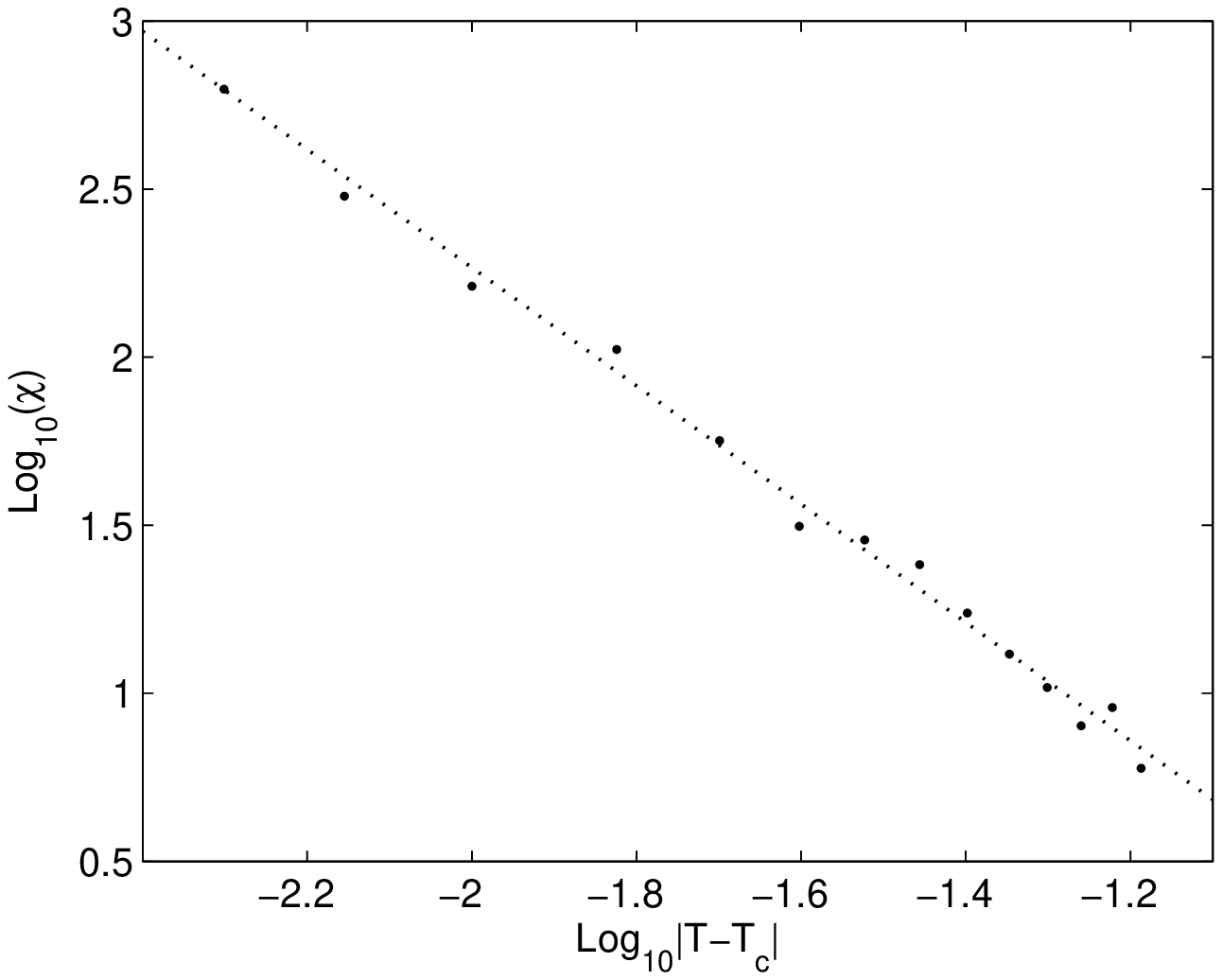}}
\caption{\label{fig:chicrit_Mezzo}Log-log scale plot of magnetic
susceptibility versus $|T-T_c|$ for a spin-1/2 system subject to
the diffusive dynamics ($\bullet$). The straight line fitting the
data has a slope $\gamma=1.761 \pm 0.049$ consistent with the
canonical critical exponent.}
\end{figure}

In Fig.~\ref{fig:mcrit_Mezzo} we plotted the data of magnetization
fitted by the power law
\begin{equation}
\label{eq:mpowerlaw} m(T) \sim |T-T_c|^{\beta}.
\end{equation}
These data were used to estimate both the transition critical
temperature and the relevant critical exponent. The estimated
values are respectively $T_c^{S=1/2} = 2.602 \pm 0.001$ and $\beta
= 0.123 \pm 0.005$. The latter is in good agreement with the
relevant critical exponent of the two-dimensional Ising model,
while the former is significantly higher than the exact one
$T_{c}^{Ising}= \frac {2 J}{\log( 1 + \sqrt{2})} \approx2.269$,
but there is a fairly good agreement with the value $T_{c}^{S=1/2}
= 2.612 \pm 0.001$ found in \cite{buonsante}.
As shown in Fig.~\ref{fig:cvcrit_Mezzo} specific heat behaves like
the function
\begin{equation}
\label{eq:cvpowerlaw} f(T) = a + b\:log(|T-T_c|)
\end{equation} which corresponds to a logarithmic divergence for the observable at the
critical temperature.
In Fig.~\ref{fig:chicrit_Mezzo} we represented a log-log scale
plot of magnetic susceptibility which is suitably fitted by a
straight line with slope $\gamma=1.761 \pm 0.049$. This means that
\begin{equation}
\label{eq:chipowerlaw} \chi(T) \sim |T-T_c|^{\gamma}.
\end{equation}

Hence, for the three critical exponents measured, there is a very
good agreement with the canonical case: $\beta^{Ising} = 1/8$,
$\alpha^{Ising} = 0$ and $\gamma^{Ising} = 7/4$.

\begin{figure}[tb]
\resizebox{0.6\columnwidth}{!}{\includegraphics{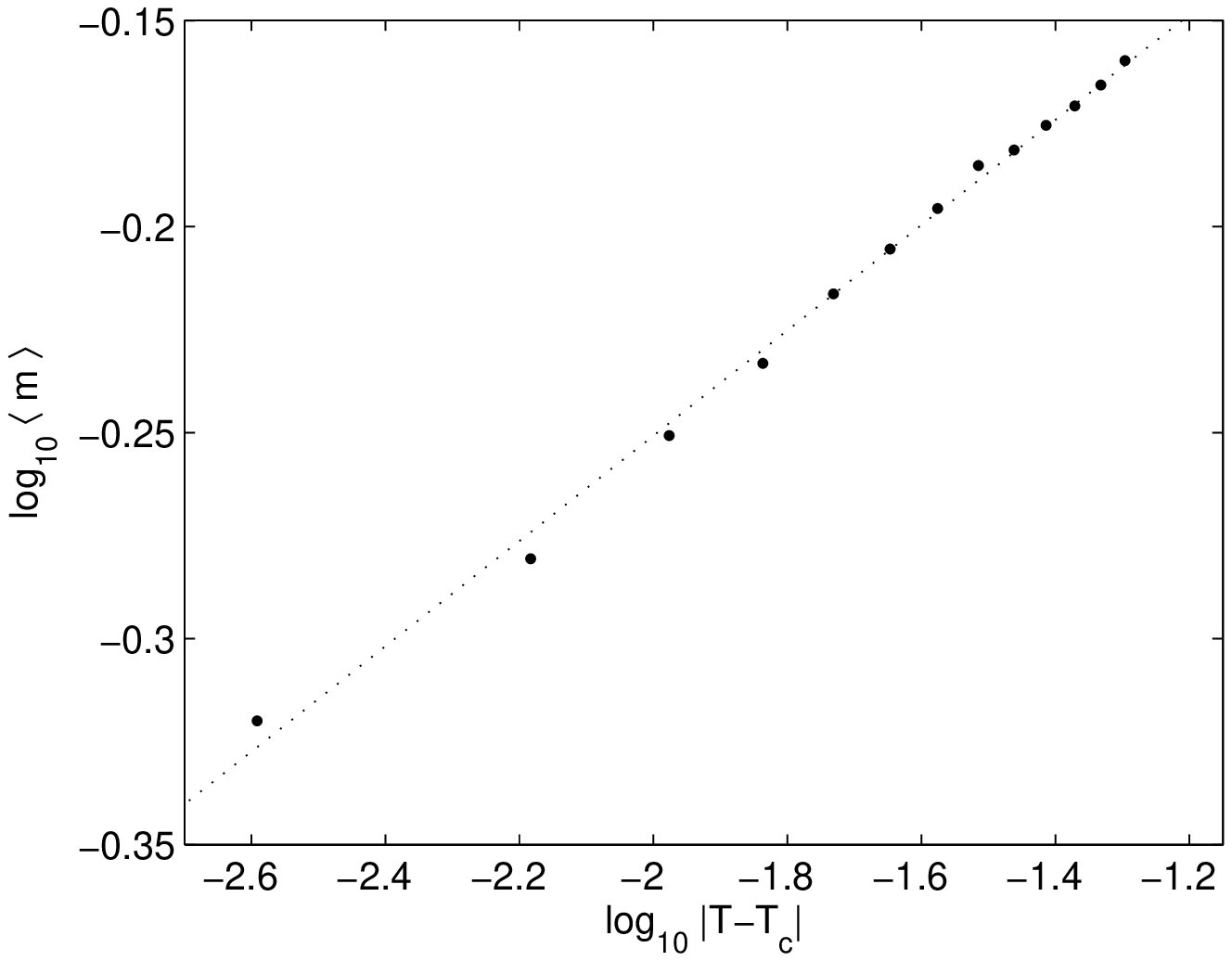}}
\caption{\label{fig:mcrit} Log-log scale plot of magnetization
versus $|T-T_c|$ for a spin-1 Ising system subject to the
diffusive dynamics ($\bullet$). The measures were performed on a
$1600 \times 1600$ array of spins. The best fit is represented by
the dotted line $y = A ~ |T-T_c|^{\beta}$. The estimated values
for the critical temperature and for the exponent are $T_c = 1.955
\pm 0.002$ and $\beta = 0.126 \pm 0.005$, respectively; the latter
is consistent with the relevant canonical one.}
\end{figure}

\begin{figure}[tb]
\resizebox{0.6\columnwidth}{!}{\includegraphics{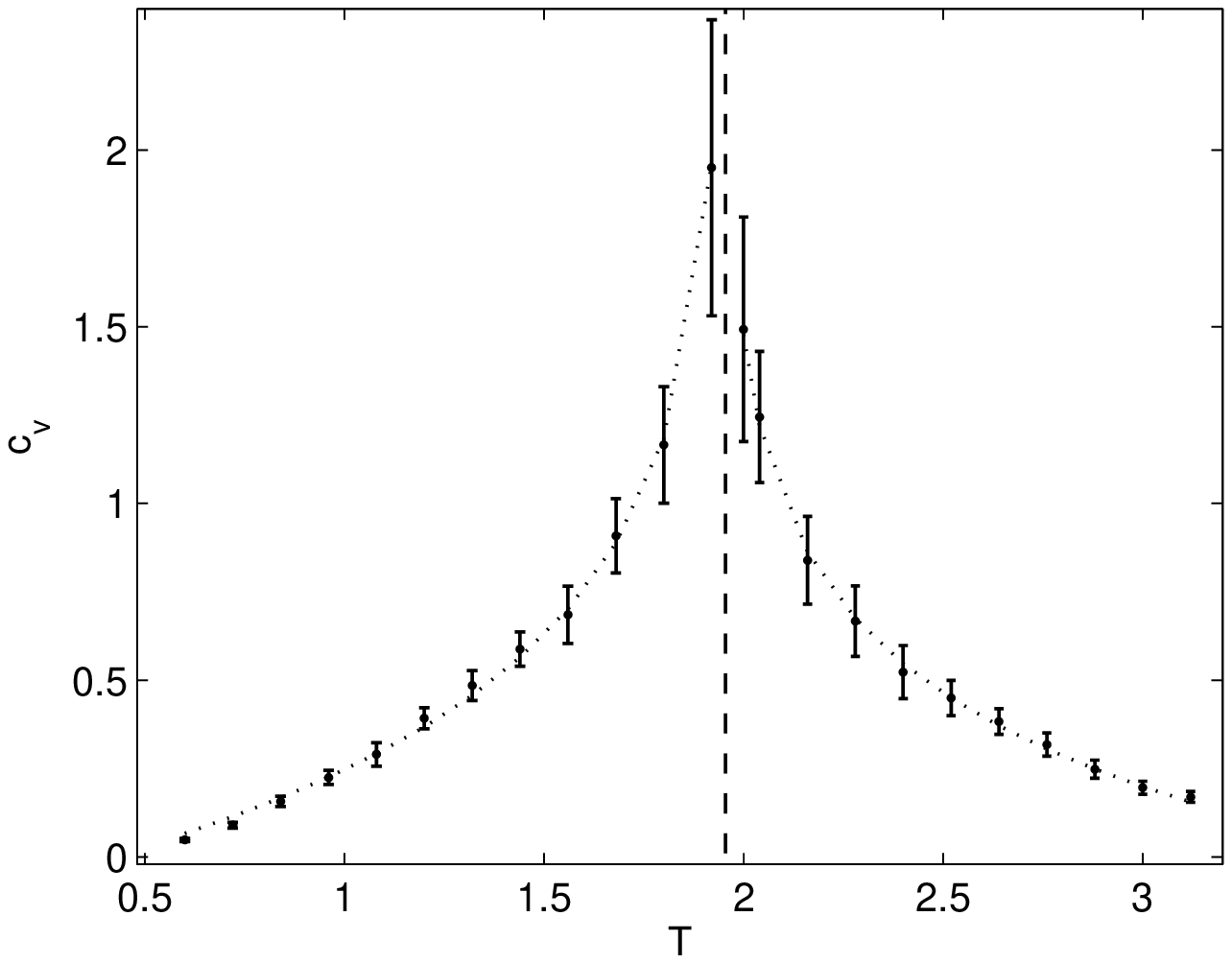}}
\caption{\label{fig:cvcrit} Specific heat for a spin-1 Ising
system subject to the diffusive dynamics ($\bullet$). The dotted
curves fitting the data are of the form $f(T) = a + b ~
log(|T-T_{c}|)$. The vertical dashed line indicates the estimated
value of the critical temperature.}
\end{figure}

\begin{figure}[tb]
\resizebox{0.6\columnwidth}{!}{\includegraphics{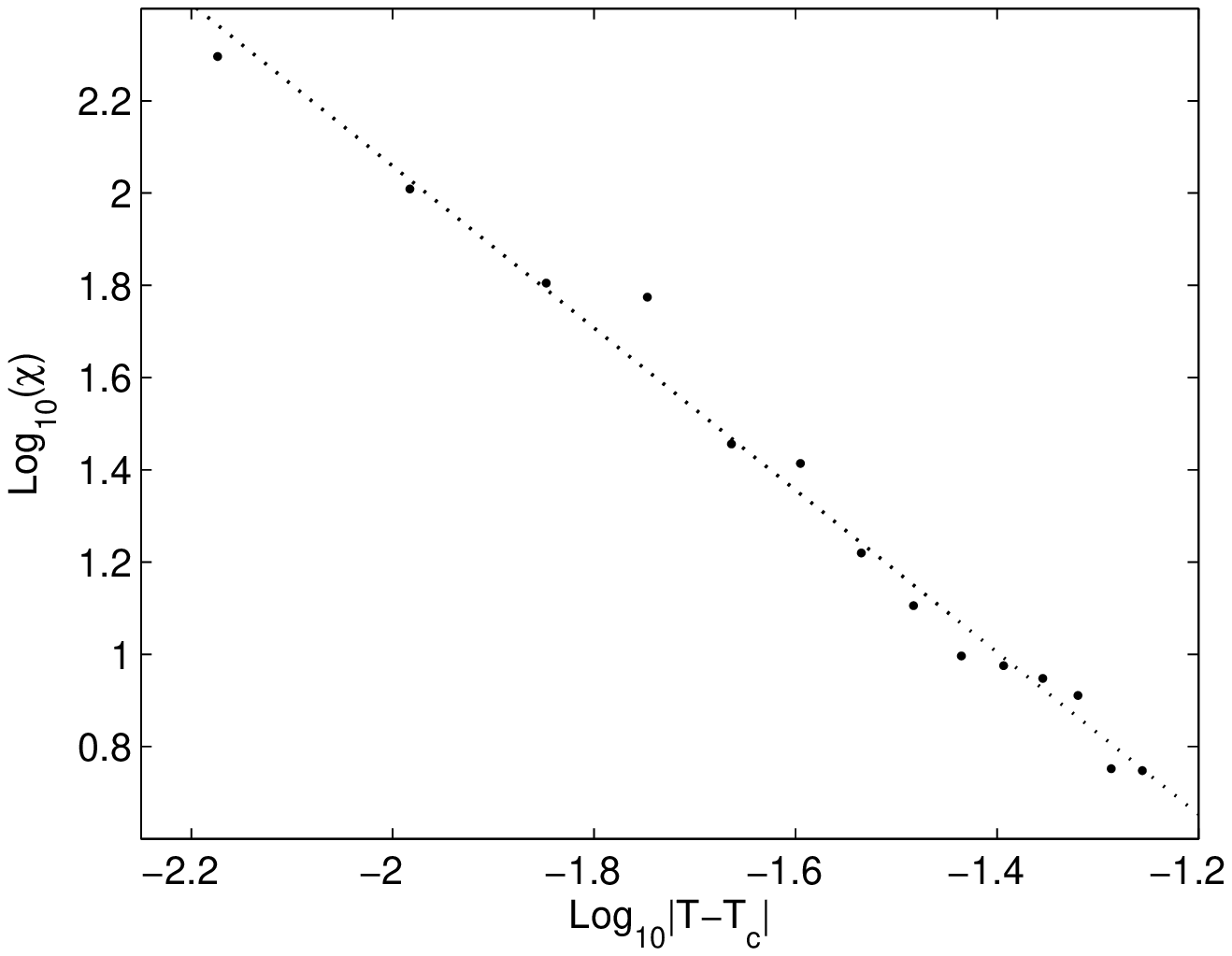}}
\caption{\label{fig:chicrit}Log-log scale plot of magnetic
susceptibility versus $|T-T_c|$ for a spin-1 system subject to the
diffusive dynamics ($\bullet$). The straight line fitting the data
has a slope $\gamma=1.756 \pm 0.064$ consistent with the canonical
critical exponent.}
\end{figure}

Now let us consider the spin-1 system: analogous results have been
gathered. Fig.~\ref{fig:mcrit} shows that magnetization data are
consistent with the same power law of Eq.~(\ref{eq:mpowerlaw})
with critical exponent $\beta = 0.126 \pm 0.005$ and critical
temperature $T_c = 1.955 \pm 0.002$ higher than values
($T_{c}^{S=1}\approx 1.695$) obtained in
\cite{hoston,berker,adler,wang,dasilva}
Also the specific heat behaves according to
Eq.~(\ref{eq:cvpowerlaw}) hence, again, a logarithmic divergence
is obtained at about T$_{c}$ (Fig.~\ref{fig:cvcrit}).
Finally, the magnetic susceptibility follows the same power law of
Eq.~(\ref{eq:chipowerlaw}) with $\gamma = 1.756 \pm 0.064$
(Fig.~\ref{fig:chicrit}).
Results explained so far point out that, different models are
similarly affected by the diffusive dynamics. In particular, the
analyzed spin-1/2 spin-1 Ising systems subject to our dynamics
share the same same universality class (which is consistent to
analytical results), despite their critical temperatures are both
15{\%} circa larger than their canonical counterparts.

As observed in \cite{buonsante}, such a quantitative difference
cannot be overcome by a simple rescaling of the temperature;
conversely, the exact critical temperature was restored by
increasing the density of walkers.

\section{Relaxation}
\label{relaxation}
In this section we deal with the properties featured by the system
when its initial configuration is paramagnetic and the temperature
is low $(T \ll T_{c})$. Of course, the time required by the walker
to lead the system to equilibrium is much larger than that needed
when the system is initialized ferromagnetic.

\begin{figure}[tb]
\resizebox{0.45\columnwidth}{!}{\includegraphics{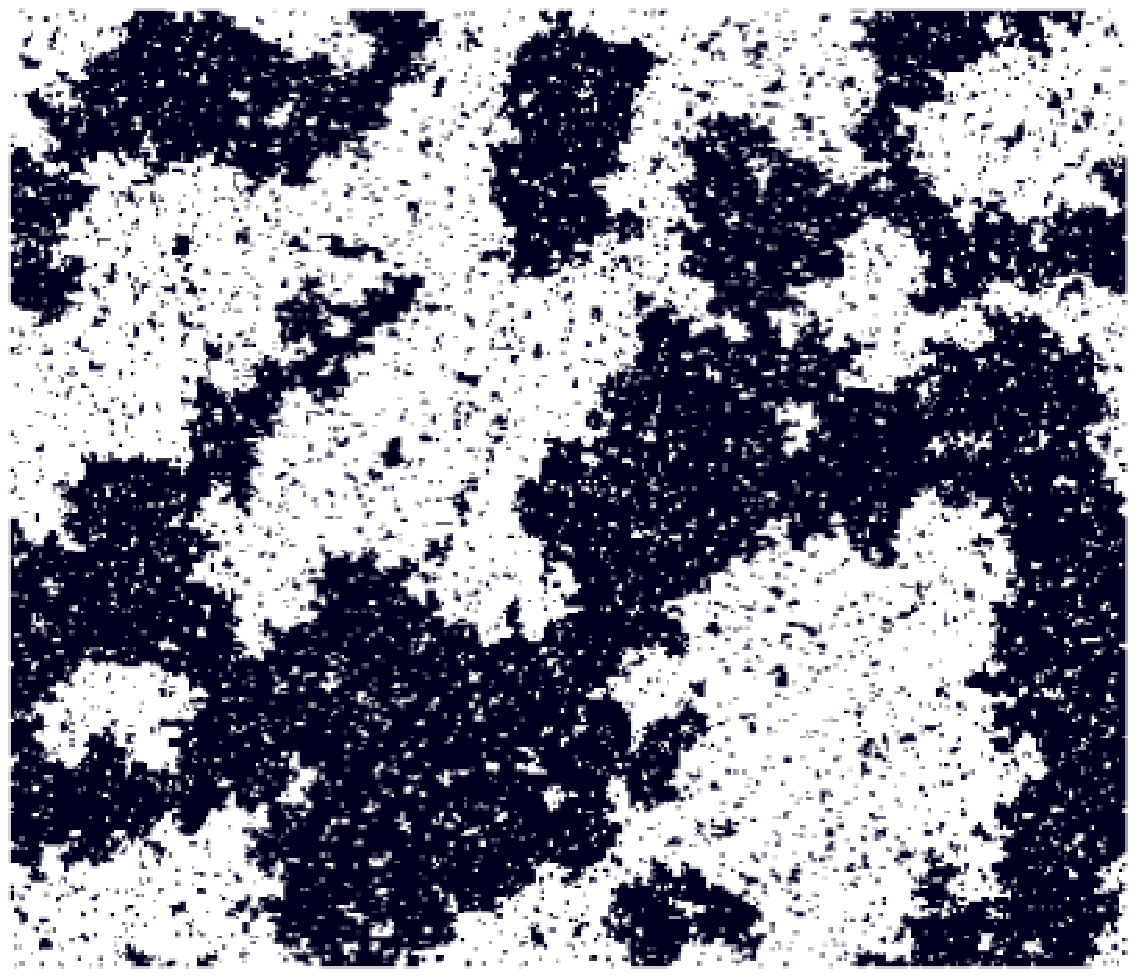}}
\resizebox{0.45\columnwidth}{!}{\includegraphics{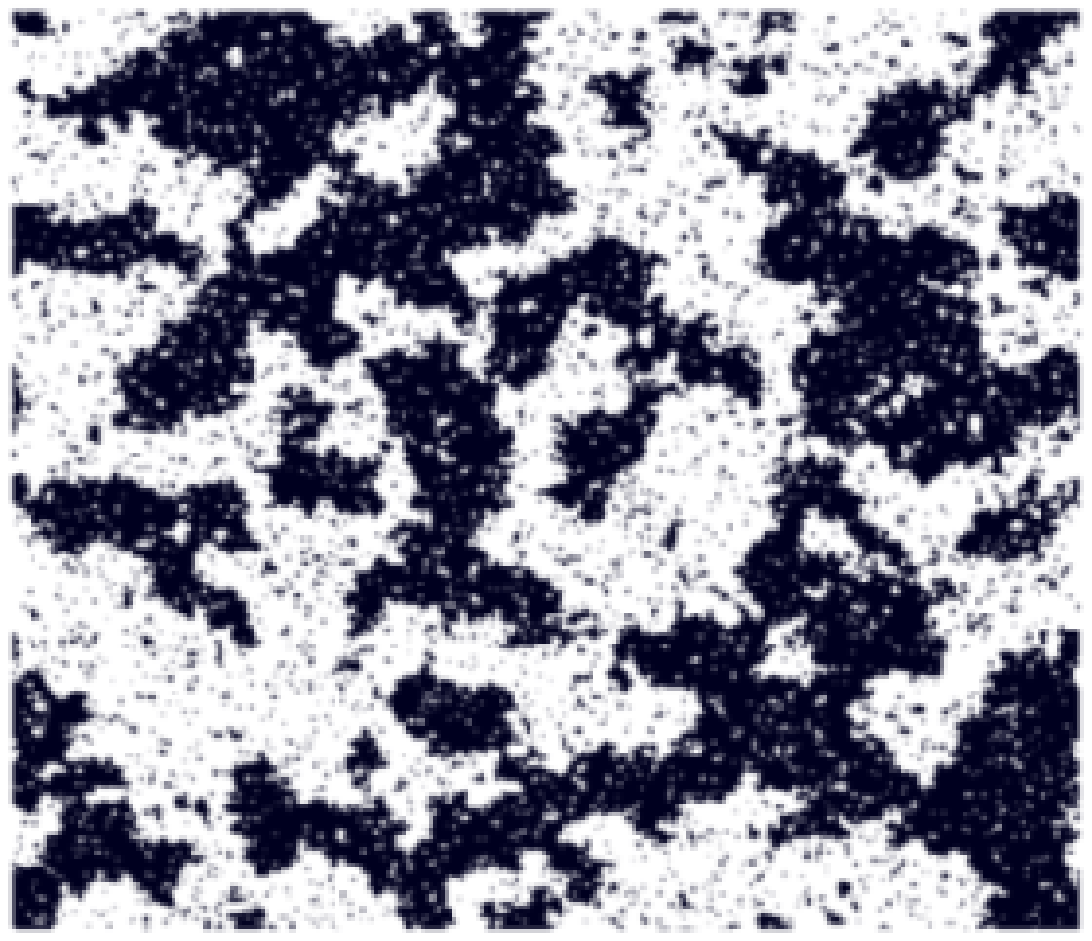}}
\caption{\label{fig:domain_Mezzo}Two snapshots showing typical
magnetic configurations for a $400 \times 400$ spin-1/2 Ising
lattice with $\langle m \rangle = 0.82$ subject to the diffusive
dynamics at $T = 2.46$ (left panel) and to the Glauber one at $T =
2.15$ (right panel). Note that the left figure shows smoother
boundaries and that the same magnetization is attained for
different temperatures.}
\end{figure}

\begin{figure}[tb]
\resizebox{0.45\columnwidth}{!}{\includegraphics{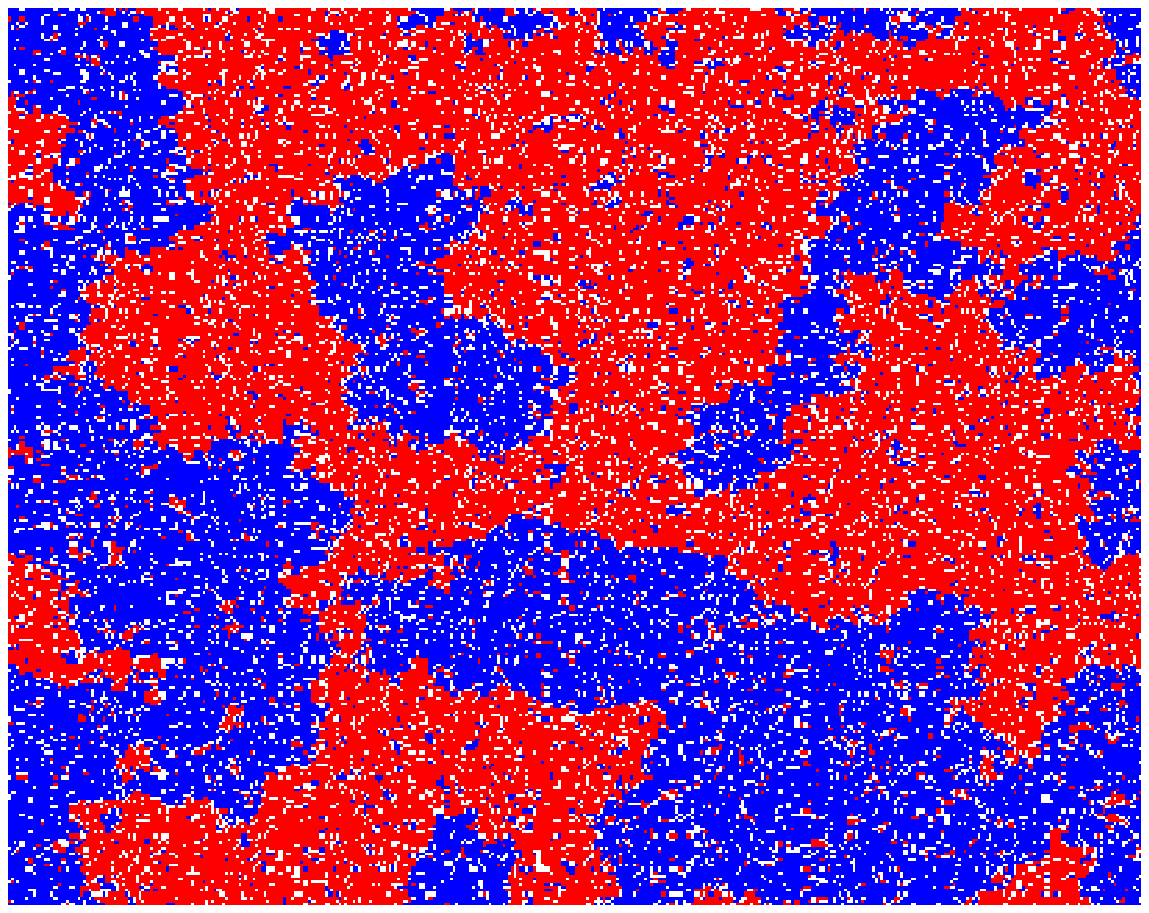}}
\resizebox{0.45\columnwidth}{!}{\includegraphics{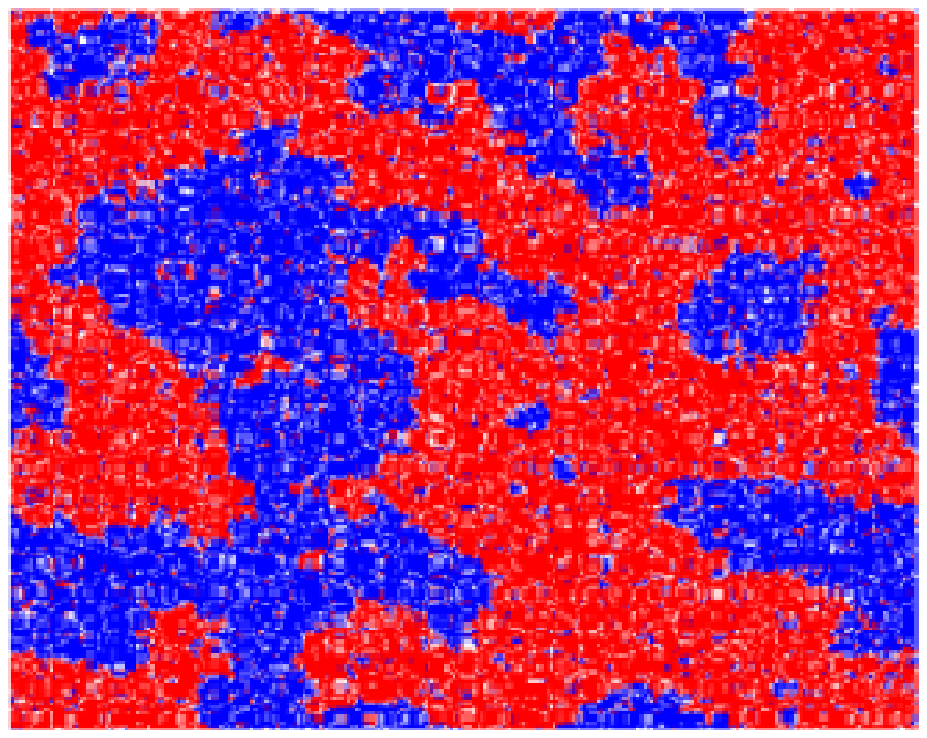}}
\caption{\label{fig:domain}(Color on line) Two snapshots showing
typical magnetic configurations for a $400 \times 400$ spin-1
Ising lattice with $\langle m \rangle = 0.78$ subject to the
diffusive dynamics at $T = 1.84$ (left panel) and to the Glauber
one at $T = 1.60$ (right panel). Null spins are colored white.
Notice that the difference between these pictures is not so marked
as that found in the previous figure.}
\end{figure}

For both spin-1/2 and spin-1 systems, starting with a low
magnetization (whatever their arrangement), we notice the
formation of domains characterized by a different orientation of
their spins. Carrying on with the simulation, one of the domains
can prevail against the others and a nearly ferromagnetic
situation is established. However, at very low temperatures, this
evolution may be delayed by the appearance of metastable states.
These states correspond to regularly shaped domains so that the
lattice appears striped. Such configurations also occur when a
non-diffusive dynamics is adopted, though less often. We also
compared the typical magnetic configurations pertaining to our
diffusive dynamics to the more traditional Metropolis dynamics,
exploiting the typewriter sequence updating. Interestingly, in the
former case, clusters display smoother boundaries, especially for
the spin-1/2 system
(Fig.~\ref{fig:domain_Mezzo},~\ref{fig:domain}). A deep study of
the geometry of these clusters will be the subject of a future
paper \cite{forth}.

Now it is worth deepening the particular role played by the null
spin in the case $S=1$. The state $\sigma = 0$ provides not only a
further option for the spin variables, but it also shows the
property of being energetically neutral. As a consequence, null
spins are not expected to form wide clusters, but rather to be
found on the boundaries between positive and negative clusters. In
particular, they are likely to stay on those sites such that
$\sum_{j=1}^{4} \sigma_i = 0$.

Finally, we note that, in the spin-1 system, the existence of a
third state makes transitions among spin states more likely to
happen. In fact, in general, when the number of states increases,
there is also a rise in the number of possible convenient events
so that a spin-flip gets more and more probable. This
consideration also provides a reason why the critical temperature
for a spin-1 system must be lower than its spin-1/2 counterpart.
An analogous consideration may also be applied to spin-S
arrangements with S$>$1 \cite{blote}.


\section{Conclusions}
\label{concl}

The new algorithm we introduced realizes, by means of a random
walker, a diffusive dynamics to be applied to an Ising
ferromagnet. Such a model provides a proper alternative to the
usual methods of updating the spin-system and it can also be
useful in order to investigate the interaction of diffusing
excitations with spins.

A fundamental feature of our algorithm is that it can be adapted
to a number of other physical systems, as it simply requires the
system to be represented by an arbitrary arrangement of sites,
each one related to a discrete variable, and to be endowed with a
proper set of local dynamics rules. Due to the arbitrariness of
the arrangement, we can consider systems implemented on general
discrete networks, ranging from completely disordered to fractal.
Moreover, as a result of our extension (see Sec.~\ref{dtd}), the
walker realizing the dynamics can deal with finite-multistate
local variables. Therefore, our algorithm can also be applied to
the q-state Potts model and, clearly, to all the physical systems
related to that model (such as lattice gas, site and bond
percolation, discrete vertex model). Finally, as far the local
dynamics rules, the equations described in Sec.~\ref{dtd}, could
be properly modified according to the particular Hamiltonian
pertaining to the system taken into account. For example, for the
Randomly Coupled Ferromagnet \cite{lemke}, a different estimate of
the energy variation consequent to a spin-flip would be reflected
by the probability of Eq.~(\ref{eq:probability}).

However, notice that, in general, the peculiar diffusive character
of the dynamics is preserved.

As far the thermodynamic of the spin systems considered in this
work, we found that the diffusive character of the dynamics leads
to a critical temperature which is significantly larger than the
canonical one, notwithstanding the universality class is
preserved. This result constitutes an interesting confirmation
that, according to the renormalization group theory, the
universality class is just concerned with the geometry of the
lattice and the symmetry of the ordered state. In fact, as
supported by our simulations, the critical exponents we measured
are the same for spin-1/2 and spin-1 Ising systems (on a squared
lattice), and, even more interestingly, they are also unaffected
by our non-canonical dynamics. Moreover, the rise in the critical
temperature and the conservation of the universality class are
very effects of the diffusive dynamics as they seem not to be due
to a particular choice of the model.

The preservation of the universality class also suggests that the
stationary state reached by the system has to be regarded as a
non-canonical equilibrium state. In addition, we recall that such
a steady state is definitely independent on the initial
conditions. On the other hand, by increasing the density of the
walkers, we expect to recover the canonical Boltzmann distribution
(as shown in \cite{buonsante}).

Of course, it would be quite interesting also to study what
happens on spin-S (S$>$1) systems or on higher dimensional
lattices.

However, what seems to be more interesting up to now is a
geometrical analysis of magnetic clusters and a characterization
of the biased random walker which will be the subject of a
forthcoming paper \cite{forth}.

%

\end{document}